\documentclass[prd,twocolumn,tightenlines]{revtex4-1}
\usepackage{tipa}
\usepackage{mathrsfs}
\usepackage{amsmath}
\usepackage{amssymb}
\usepackage{graphicx}

\newcommand{\be}{\begin{eqnarray}}
\newcommand{\ee}{\end{eqnarray}}

 \newcommand{\gsim}{\mathrel{\hbox{\rlap{\lower.55ex \hbox {$\sim$}}
                   \kern-.3em \raise.4ex \hbox{$>$}}}}
\newcommand{\lsim}{\mathrel{\hbox{\rlap{\lower.55ex \hbox {$\sim$}}
                   \kern-.3em \raise.4ex \hbox{$<$}}}}

\begin{document}

\title{ Heavy quarkonium photoproduction in ultrarelativistic heavy ion collisions }

\author{Gong-Ming Yu$^{1}$}
\email{ygmanan@163.com}

\author{Yan-Bing Cai$^{2}$}
\email{myparticle@163.com}

\author{Yun-De Li$^{2}$}
\email{yndxlyd@163.com}

\author{Jian-Song Wang$^{1,}$}
\email{jswang@impcas.ac.cn}

\address{ $^{1}$Institute of Modern Physics, Chinese Academy of Sciences, Lanzhou 730000, China\\$^{2}$Department of Physics, Yunnan University, Kunming 650091, China }





\begin{abstract}
Based on the factorization formalism of non-relativistic Quantum
Chromodynamics (NRQCD), we calculate the production cross section for the charmonium
($J/\psi$, $\psi(2S)$, $\chi_{cJ}$, $\eta_{c}$, and $h_{c}$) and the
bottomonium ($\Upsilon(nS)$, $\chi_{bJ}$, $\eta_{b}$, and $h_{b}$)
produced by the hard photoproduction processes and fragmentation
processes in relativistic heavy ion collisions. It is shown that the existing experimental data on
heavy quarkonium production at Large Hadron Collider (LHC) can be
described in the framework of the NRQCD formalism, and the phenomenological value of
matrix element for color-singlet and color-octet components give
main contribution. The numerical results of photoproduction
processes and fragmentation processes for the heavy quarkonium production become prominent in p-p collisions and Pb-Pb collisions at Large Hadron Collider (LHC)
energies.


\end{abstract}

\maketitle



\section{Introduction}

Hadronic processes for large transverse momentum heavy quarkonium is
a vital goal in ultrarelativistic heavy ion collisions at Large
Hadron Collider (LHC). In the last years, the measurements reported
by the ALICE collaboration\cite{1,2,3,4,5}, CMS
collaboration\cite{6,7,8,9}, ATLAS collaboration\cite{10,11}, and
LHCb collaboration\cite{12,13,14} at the Large Hadron Collider (LHC)
have posed significant challenges to our understanding of the heavy
quarkonium production.  Furthermore, several theoretical approaches
have been proposed for the calculation of these states, as for
instance, the color-singlet mechanism\cite{15,16}, the color-octet
mechanism\cite{17,18,19,20,21,22}, the color evaporation
mechanism\cite{23,24}, the color-dipole mechanism\cite{25,26,27,28},
the mixed heavy-quark hybrids mechanism\cite{29}, the recombination
mechanism\cite{30,31,32}, the photoproduction
mechanism\cite{33,34,35,36,37,38,39,40,41,42,43,44,45,46,47,48,49},
the potential non-relativistic Quantum Chromodynamics (pNRQCD)
approach\cite{50,51,52}, the transverse-momentum-dependent
factorization approach\cite{53}, the transport
approach\cite{54,55,56}, the $k_{T}$-factorization
approach\cite{57,58,59,60,61}, the fragmentation
approach\cite{62,63,64,65,66,67,68,69,70,71,72,73,74}, and the
non-relativistic Quantum Chromodynamics (NRQCD)
approach\cite{75,76,77,78,79,80,81,82,83,84,85,86,87,88,89,90,91,92,93,94,95}.
Among them, the non-relativistic Quantum Chromodynamics (NRQCD)
approach is the most successful in phenomenological study. According
to the non-relativistic Quantum Chromodynamics (NRQCD) in addition
to color-singlet component one has to take into account
contributions of color-octet components with the nonperturbative
long-distance matrix elements. This formalism implies a separation
of short-distance coefficients, which can be calculated
perturbatively as expansions in the running coupling constant
$\alpha_{s}$, from long-distance matrix elements, which can be
extracted from experiment. The long-distance matrix elements are
process-independent, and can be classified in terms of their scaling
in $v$, the relative velocity of the heavy quarks in the bound
state. A crucial feature for this formalism is that it takes into
account the $Q\bar{Q}$ Fock space\cite{77}, which is spanned by the
states $|n\rangle=|^{2S+1}L_{J}^{(c)}\rangle$ with definite spin
$S$, orbital angular momentum $L$, total angular momentum $J$, and
color multiplicity $c$. In particular, this formalism predicts the
existence of color-octet processes by the nonperturbative emission
of soft gluons in nature.

Although considerable efforts both in experiment and theory, the
heavy quarkonium production mechanism is still not fully understood.
In the present work, we extend the hard photoproduction
mechanism\cite{96,97,98} which plays a fundamental role in the
electron-proton deep inelastic scattering at the Hadron Electron
Ring Accelerator to the heavy quarkonium production in p-p
collisions and Pb-Pb collisions at Large Hadron Collider (LHC)
energies. At high energies, the nucleus (the charged partons) can
emit high-energy photons (and hadron-like photons) in
ultrarelativistic nucleus-nucleus collisions. The hard
photoproduction processes may be direct and resolved that are
sensitive to the gluon distribution in the nucleus. In the hard
direct photoproduction processes, the high-energy photon emitted
from the nucleus (the charged parton of the incident nucleus)
interacts with the parton of another incident nucleus. In the hard
resolved photoproduction processes, the uncertainty principle allows
the high-energy hadronlike photon emitted from the nucleus (the
charged parton of the incident nucleus) for a short time to
fluctuate into a quark-antiquark pair which then interacts with the
partons of another incident nucleus. For the fragmentation
processes, the jets can produced by the hard photoproduction
processes and the hard resolved photoproduction processes.
Subsequently the jets will fragment into heavy quarkonium.

The paper is organized as follows. In Sec.II we present the
photoproduction of heavy quarkonium at Large Hadron Collider (LHC)
energies. The production rate for fragmentation processes is also
discussed. The numerical results for large-$p_{T}$ heavy quarkonium
in p-p collisions and Pb-Pb collisions at Large Hadron Collider
(LHC) energies are plotted in Sec.III. Finally, the conclusion is
given in Sec.IV.


\section{Photoproduction processes for large-$P_{T}$ heavy quarkonium}

The non-relativistic Quantum Chromodynamics (NRQCD)
approach\cite{99} which is introduced in 1995 has become the
standard framework to study heavy-quarkonium physics. In the
non-relativistic Quantum Chromodynamics (NRQCD), the Fock state
structure of heavy quarkonium state is\cite{99,100}
\begin{eqnarray}\label{y1}
|^{2S+1}L_{J}\rangle&=&\mathcal
{O}(1)|^{2S+1}L_{J}^{[1]}\rangle\nonumber\\[1mm]
&&+\mathcal {O}(v)|^{2S+1}(L\pm1)_{J'}^{[8]}g\rangle\nonumber\\[1mm]
&&+\mathcal {O}(v^{2})|^{2(S\pm1)+1}(L\pm1)_{J'}^{[8]}g\rangle\nonumber\\[1mm]
&&+\mathcal {O}(v^{2})|^{2S+1}L_{J}^{[1,8]}gg\rangle\nonumber\\[1mm]
&&+\cdots\cdots,
\end{eqnarray}
where $v$ is the relative velocity of the heavy quarks in heavy
quarkonium.

\begin{figure*}
\includegraphics[width=15cm,height=10cm]{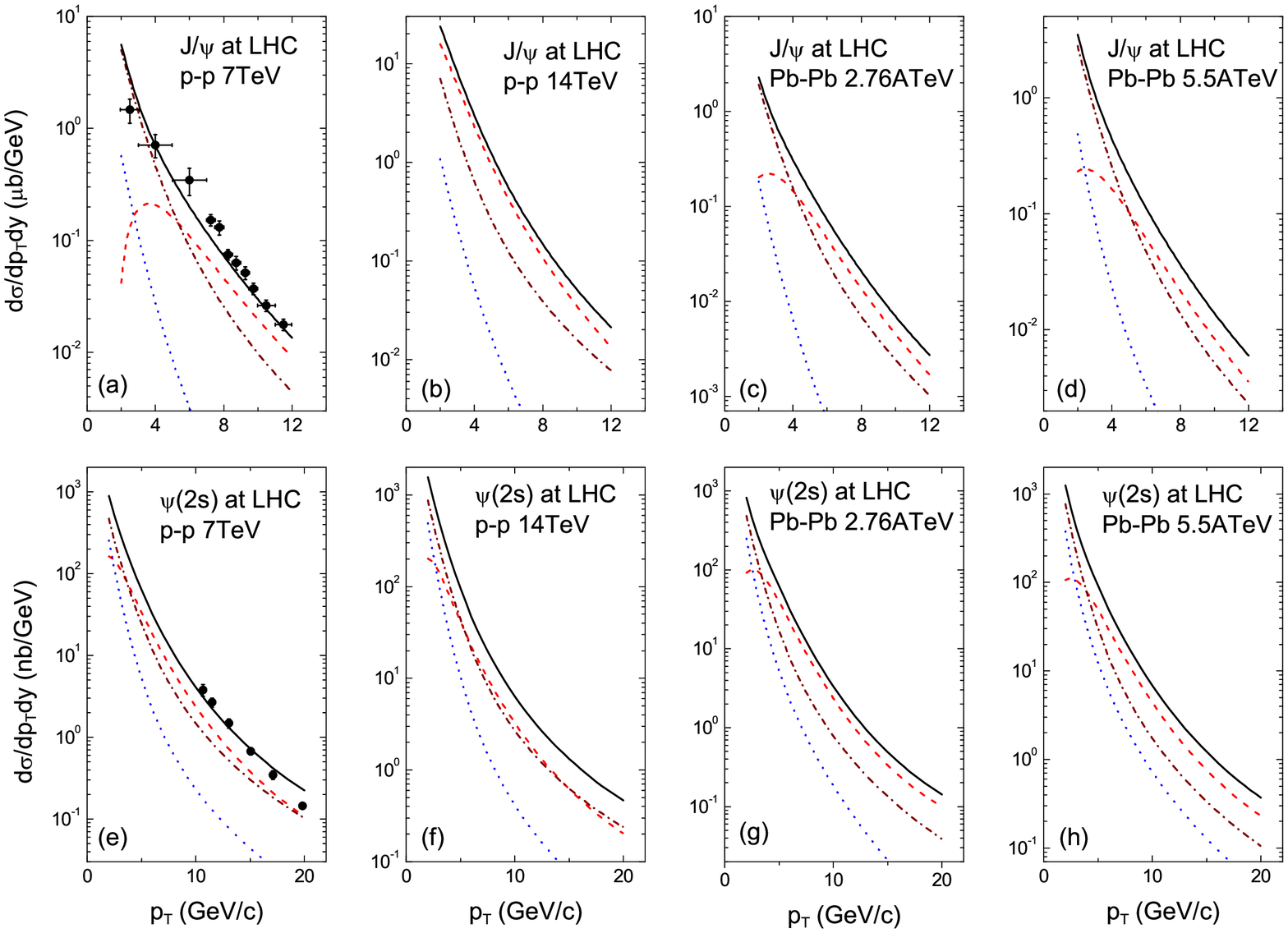}
\caption{\label{T1}  The invariant cross section of large-$p_{T}$
$J/\psi$ and $\psi(2S)$ production for p-p collisions ($\sqrt{s}=7.0
TeV$ and $\sqrt{s}=14.0 TeV$) and Pb-Pb collisions ($\sqrt{s}=2.76
TeV$ and $\sqrt{s}=5.5 TeV$) at LHC. The dashed line (red line) for
the initial parton hard scattering processes (LO+LO-fra.), the dotted line (blue
line) for the semielastic photoproduction processes (semi.(dir.+res.)+semi.(dir.+res.)-fra.), the
dashed-dotted line (wine line) for the inelastic photoproduction
processes (inel.(dir.+res.)+inel.(dir.+res.)-fra.), and the solid line (black) is for the sum of
the above processes. The $J/\psi$ data points are from the ALICE
Collaboration\cite{1}, and the $\psi(2S)$ data points are from the
CMS Collaboration\cite{8}.}
\end{figure*}

The wavefunctions of the charmonium ($J/\psi$, $\psi(2s)$,
$\chi_{cJ}$, $\eta_{c}$, and $h_{c}$) can be written as
\begin{eqnarray}\label{y2}
|J/\psi\rangle&=&\langle\mathcal
{O}^{J/\psi}[^{3}S_{1}^{[1]}]\rangle|Q\bar{Q}[^{3}S_{1}^{[1]}]\rangle\nonumber\\[1mm]
&&+\langle\mathcal
{O}^{J/\psi}[^{1}S_{0}^{[8]}]\rangle|Q\bar{Q}[^{1}S_{0}^{[8]}]g\rangle\nonumber\\[1mm]
&&+\langle\mathcal
{O}^{J/\psi}[^{3}S_{1}^{[8]}]\rangle|Q\bar{Q}[^{3}S_{1}^{[8]}]gg\rangle\nonumber\\[1mm]
&&+\sum_{J'}\langle\mathcal
{O}^{J/\psi}[^{3}P_{J'}^{[8]}]\rangle|Q\bar{Q}[^{3}P_{J'}^{[8]}]g\rangle\nonumber\\[1mm]
&&+\cdots\cdots,
\end{eqnarray}
\begin{eqnarray}\label{y3}
|\psi(2S)\rangle&=&\langle\mathcal
{O}^{\psi(2S)}[^{3}S_{1}^{[1]}]\rangle|Q\bar{Q}[^{3}S_{1}^{[1]}]\rangle\nonumber\\[1mm]
&&+\langle\mathcal
{O}^{\psi(2S)}[^{1}S_{0}^{[8]}]\rangle|Q\bar{Q}[^{1}S_{0}^{[8]}]g\rangle\nonumber\\[1mm]
&&+\langle\mathcal
{O}^{\psi(2S)}[^{3}S_{1}^{[8]}]\rangle|Q\bar{Q}[^{3}S_{1}^{[8]}]gg\rangle\nonumber\\[1mm]
&&+\sum_{J'}\langle\mathcal
{O}^{\psi(2S)}[^{3}P_{J'}^{[8]}]\rangle|Q\bar{Q}[^{3}P_{J'}^{[8]}]g\rangle\nonumber\\[1mm]
&&+\cdots\cdots,
\end{eqnarray}
\begin{eqnarray}\label{y4}
|\chi_{cJ}\rangle&=&\langle\mathcal
{O}^{\chi_{cJ}}[^{3}P_{J}^{[1]}]\rangle|Q\bar{Q}[^{3}P_{J}^{[1]}]\rangle\nonumber\\[1mm]
&&+\langle\mathcal
{O}^{\chi_{cJ}}[^{1}S_{0}^{[8]}]\rangle|Q\bar{Q}[^{1}S_{o}^{[8]}]g\rangle\nonumber\\[1mm]
&&+\langle\mathcal
{O}^{\chi_{cJ}}[^{3}S_{1}^{[8]}]\rangle|Q\bar{Q}[^{3}S_{1}^{[8]}]g\rangle\nonumber\\[1mm]
&&+\langle\mathcal
{O}^{\chi_{cJ}}[^{1}P_{1}^{[8]}]\rangle|Q\bar{Q}[^{1}P_{1}^{[8]}]g\rangle\nonumber\\[1mm]
&&+\langle\mathcal
{O}^{\chi_{cJ}}[^{3}P_{J}^{[8]}]\rangle|Q\bar{Q}[^{3}P_{J}^{[8]}]gg\rangle\nonumber\\[1mm]
&&+\langle\mathcal
{O}^{\chi_{cJ}}[^{3}D_{J}^{[8]}]\rangle|Q\bar{Q}[^{3}D_{J}^{[8]}]g\rangle\nonumber\\[1mm]
&&+\cdots\cdots,
\end{eqnarray}
\begin{eqnarray}\label{y5}
|\eta_{c}\rangle&=&\langle\mathcal
{O}^{\eta_{c}}[^{1}S_{0}^{[1]}]\rangle|Q\bar{Q}[^{1}S_{0}^{[1]}]\rangle\nonumber\\[1mm]
&&+\langle\mathcal
{O}^{\eta_{c}}[^{1}S_{0}^{[8]}]\rangle|Q\bar{Q}[^{1}S_{0}^{[8]}]g\rangle\nonumber\\[1mm]
&&+\langle\mathcal
{O}^{\eta_{c}}[^{3}S_{1}^{[8]}]\rangle|Q\bar{Q}[^{3}S_{1}^{[8]}]g\rangle\nonumber\\[1mm]
&&+\langle\mathcal
{O}^{\eta_{c}}[^{1}P_{1}^{[8]}]\rangle|Q\bar{Q}[^{1}P_{1}^{[8]}]g\rangle\nonumber\\[1mm]
&&+\cdots\cdots,
\end{eqnarray}
\begin{eqnarray}\label{y6}
|h_{c}\rangle&=&\langle\mathcal
{O}^{h_{c}}[^{1}P_{1}^{[1]}]\rangle|Q\bar{Q}[^{1}P_{1}^{[1]}]g\rangle\nonumber\\[1mm]
&&+\langle\mathcal
{O}^{h_{c}}[^{1}S_{0}^{[8]}]\rangle|Q\bar{Q}[^{1}S_{0}^{[8]}]g\rangle\nonumber\\[1mm]
&&+\cdots\cdots,
\end{eqnarray}
where $\langle\mathcal {O}^{H}[^{2S+1}L_{J}^{[1,8]}]\rangle$ is the
long-distance matrix element for the
charmonium\cite{18,19,90,91,92,93,94,101,102}, and  the
long-distance matrix elements for the $J/\psi$ meson are given by
\begin{eqnarray}\label{y7}
&&\langle\mathcal
{O}^{J/\psi}[^{3}S_{1}^{[1]}]\rangle=1.2GeV^{3},\nonumber\\[1mm]
&&\langle\mathcal
{O}^{J/\psi}[^{1}S_{0}^{[8]}]\rangle=(0.0180\pm0.0087)GeV^{3},\nonumber\\[1mm]
&&\langle\mathcal
{O}^{J/\psi}[^{3}S_{1}^{[8]}]\rangle=(0.0013\pm0.0013)GeV^{3},\nonumber\\[1mm]
&&\langle\mathcal
{O}^{J/\psi}[^{3}P_{0}^{[8]}]\rangle=(0.0180\pm0.0087)m_{c}^{2}\cdot
GeV^{3},\nonumber\\[1mm]
&&\langle\mathcal
{O}^{J/\psi}[^{3}P_{1}^{[8]}]\rangle=3\times(0.0180\pm0.0087)m_{c}^{2}\cdot
GeV^{3},\nonumber\\[1mm]
&&\langle\mathcal
{O}^{J/\psi}[^{3}P_{2}^{[8]}]\rangle=5\times(0.0180\pm0.0087)m_{c}^{2}\cdot
GeV^{3},
\end{eqnarray}
for $\psi(2S)$ meson,
\begin{eqnarray}\label{y8}
&&\!\!\!\!\langle\mathcal
{O}^{\psi(2S)}[^{3}S_{1}^{[1]}]\rangle=0.76GeV^{3},\nonumber\\[1mm]
&&\!\!\!\!\langle\mathcal
{O}^{\psi(2S)}[^{1}S_{0}^{[8]}]\rangle=(0.0080\pm0.0067)GeV^{3},\nonumber\\[1mm]
&&\!\!\!\!\langle\mathcal
{O}^{\psi(2S)}[^{3}S_{1}^{[8]}]\rangle=(0.00330\pm0.00021)GeV^{3},\nonumber\\[1mm]
&&\!\!\!\!\langle\mathcal
{O}^{\psi(2S)}[^{3}P_{0}^{[8]}]\rangle=(0.0080\pm0.0067)m_{c}^{2}\cdot
GeV^{3},\nonumber\\[1mm]
&&\!\!\!\!\langle\mathcal
{O}^{\psi(2S)}[^{3}P_{1}^{[8]}]\rangle=3\times(0.0080\pm0.0067)m_{c}^{2}\cdot
GeV^{3},\nonumber\\[1mm]
&&\!\!\!\!\langle\mathcal
{O}^{\psi(2S)}[^{3}P_{2}^{[8]}]\rangle=5\times(0.0080\pm0.0067)m_{c}^{2}\cdot
GeV^{3},
\end{eqnarray}
for $\chi_{cJ}$ meson,
\begin{eqnarray}\label{y9}
&&\langle\mathcal
{O}^{\chi_{c0}}[^{3}P_{0}^{[1]}]\rangle=0.054m_{c}^{2}\cdot GeV^{3},\nonumber\\[1mm]
&&\langle\mathcal {O}^{\chi_{c0}}[^{3}S_{1}^{[8]}]\rangle=(0.00187\pm0.00025)GeV^{3},\nonumber\\[1mm]
&&\langle\mathcal {O}^{\chi_{c1}}[^{3}P_{1}^{[1]}]\rangle=3\times0.054m_{c}^{2}\cdot GeV^{3},\nonumber\\[1mm]
&&\langle\mathcal {O}^{\chi_{c1}}[^{3}S_{1}^{[8]}]\rangle=3\times(0.00187\pm0.00025)GeV^{3},\nonumber\\[1mm]
&&\langle\mathcal {O}^{\chi_{c2}}[^{3}P_{1}^{[1]}]\rangle=5\times0.054m_{c}^{2}\cdot GeV^{3},\nonumber\\[1mm]
&&\langle\mathcal
{O}^{\chi_{c2}}[^{3}S_{1}^{[8]}]\rangle=5\times(0.00187\pm0.00025)GeV^{3},
\end{eqnarray}
for $\eta_{c}$ meson,
\begin{eqnarray}\label{y10}
&&\!\!\!\!\langle\mathcal
{O}^{\eta_{c}}[^{1}S_{0}^{[1]}]\rangle=\frac{1}{3}\times1.2GeV^{3},\nonumber\\[1mm]
&&\!\!\!\!\langle\mathcal
{O}^{\eta_{c}}[^{1}S_{0}^{[8]}]\rangle=\frac{1}{3}\times(0.0013\pm0.0013)GeV^{3},\nonumber\\[1mm]
&&\!\!\!\!\langle\mathcal
{O}^{\eta_{c}}[^{3}S_{1}^{[8]}]\rangle=(0.0180\pm0.0087)GeV^{3},\nonumber\\[1mm]
&&\!\!\!\!\langle\mathcal
{O}^{\eta_{c}}[^{1}P_{1}^{[8]}]\rangle=3\times(0.0180\pm0.0087)m_{c}^{2}\cdot
GeV^{3},
\end{eqnarray}
and for $h_{c}$ meson,
\begin{eqnarray}\label{y11}
&&\langle\mathcal
{O}^{h_{c}}[^{1}P_{1}^{[1]}]\rangle=3\times0.54m_{c}^{2}\cdot
GeV^{3},\nonumber\\[1mm]
&&\langle\mathcal
{O}^{h_{c}}[^{1}S_{0}^{[8]}]\rangle=3\times(0.00187\pm0.00025)GeV^{3},
\end{eqnarray}
here $m_{c}$ is the charm quark mass.

\begin{figure*}
\includegraphics[width=15cm,height=10cm]{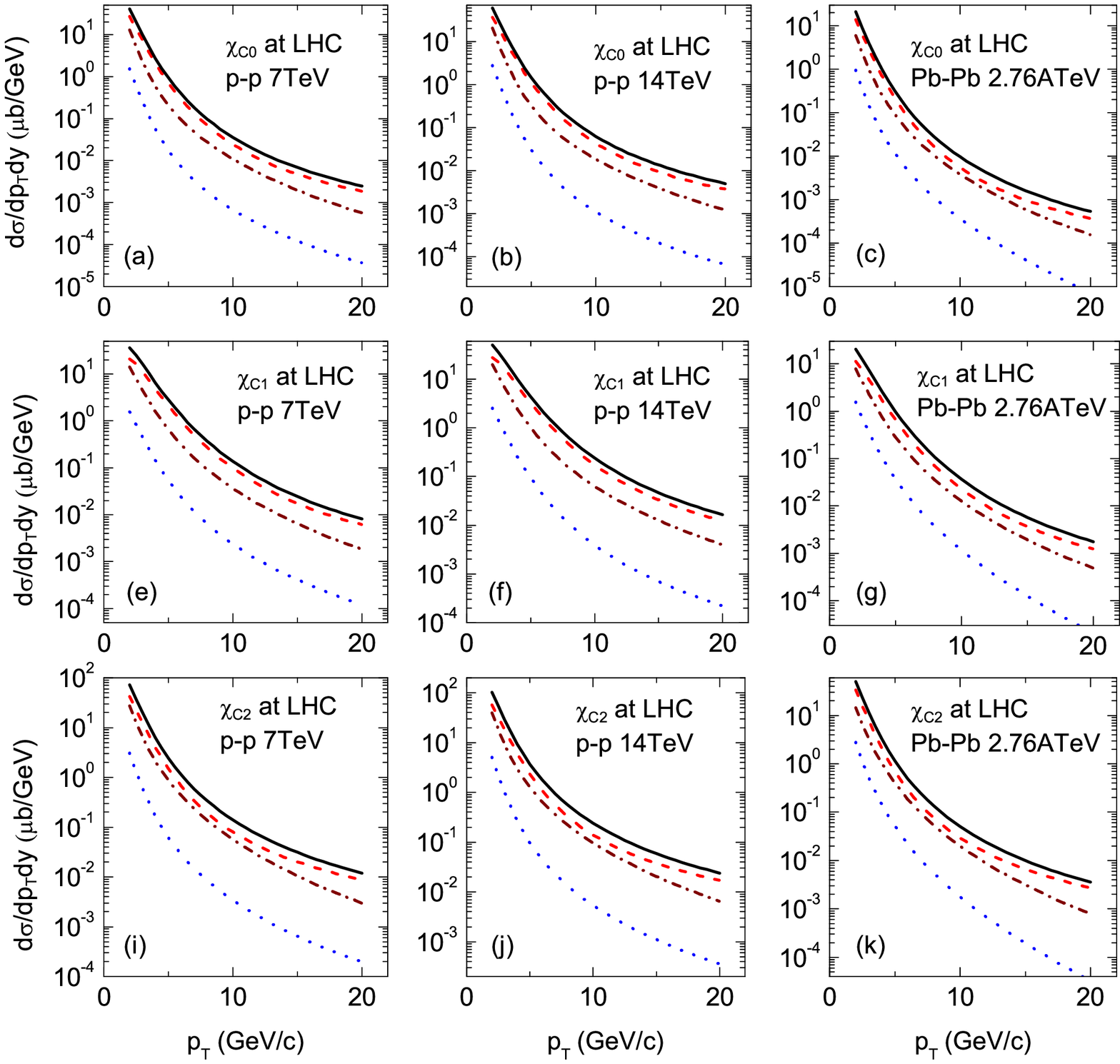}
\caption{\label{T2}  The same as Fig.\,\ref{T1} but for $\chi_{cJ}$
production in p-p collisions ($\sqrt{s}=7.0 TeV$ and $\sqrt{s}=14.0
TeV$) and Pb-Pb collisions ($\sqrt{s}=2.76 TeV$ and $\sqrt{s}=5.5
TeV$) at LHC.}
\end{figure*}

The wavefunctions of the bottomonium ($\Upsilon(ns)$, $\chi_{bJ}$,
$\eta_{b}$, and $h_{b}$) can be written as
\begin{eqnarray}\label{y12}
|\Upsilon(nS)\rangle&=&\langle\mathcal
{O}^{\Upsilon(nS)}[^{3}S_{1}^{[1]}]\rangle|Q\bar{Q}[^{3}S_{1}^{[1]}]\rangle\nonumber\\[1mm]
&&+\langle\mathcal
{O}^{\Upsilon(nS)}[^{1}S_{0}^{[8]}]\rangle|Q\bar{Q}[^{1}S_{0}^{[8]}]g\rangle\nonumber\\[1mm]
&&+\langle\mathcal
{O}^{\Upsilon(nS)}[^{3}S_{1}^{[8]}]\rangle|Q\bar{Q}[^{3}S_{1}^{[8]}]gg\rangle\nonumber\\[1mm]
&&+\sum_{J'}\langle\mathcal
{O}^{\Upsilon(nS)}[^{3}P_{J'}^{[8]}]\rangle|Q\bar{Q}[^{3}P_{J'}^{[8]}]g\rangle\nonumber\\[1mm]
&&+\cdots\cdots,
\end{eqnarray}
\begin{eqnarray}\label{y13}
|\chi_{bJ}\rangle&=&\langle\mathcal
{O}^{\chi_{bJ}}[^{3}P_{J}^{[1]}]\rangle|Q\bar{Q}[^{3}P_{J}^{[1]}]\rangle\nonumber\\[1mm]
&&+\langle\mathcal
{O}^{\chi_{bJ}}[^{1}S_{0}^{[8]}]\rangle|Q\bar{Q}[^{1}S_{o}^{[8]}]g\rangle\nonumber\\[1mm]
&&+\langle\mathcal
{O}^{\chi_{bJ}}[^{3}S_{1}^{[8]}]\rangle|Q\bar{Q}[^{3}S_{1}^{[8]}]g\rangle\nonumber\\[1mm]
&&+\langle\mathcal
{O}^{\chi_{bJ}}[^{1}P_{1}^{[8]}]\rangle|Q\bar{Q}[^{1}P_{1}^{[8]}]g\rangle\nonumber\\[1mm]
&&+\langle\mathcal
{O}^{\chi_{bJ}}[^{3}P_{J}^{[8]}]\rangle|Q\bar{Q}[^{3}P_{J}^{[8]}]gg\rangle\nonumber\\[1mm]
&&+\langle\mathcal
{O}^{\chi_{bJ}}[^{3}D_{J}^{[8]}]\rangle|Q\bar{Q}[^{3}D_{J}^{[8]}]g\rangle\nonumber\\[1mm]
&&+\cdots\cdots,
\end{eqnarray}
\begin{eqnarray}\label{y14}
|\eta_{b}\rangle&=&\langle\mathcal
{O}^{\eta_{b}}[^{1}S_{0}^{[1]}]\rangle|Q\bar{Q}[^{1}S_{0}^{[1]}]\rangle\nonumber\\[1mm]
&&+\langle\mathcal
{O}^{\eta_{b}}[^{1}S_{0}^{[8]}]\rangle|Q\bar{Q}[^{1}S_{0}^{[8]}]g\rangle\nonumber\\[1mm]
&&+\langle\mathcal
{O}^{\eta_{b}}[^{3}S_{1}^{[8]}]\rangle|Q\bar{Q}[^{3}S_{1}^{[8]}]g\rangle\nonumber\\[1mm]
&&+\langle\mathcal
{O}^{\eta_{b}}[^{1}P_{1}^{[8]}]\rangle|Q\bar{Q}[^{1}P_{1}^{[8]}]g\rangle\nonumber\\[1mm]
&&+\cdots\cdots,
\end{eqnarray}
\begin{eqnarray}\label{y15}
|h_{b}\rangle&=&\langle\mathcal
{O}^{h_{b}}[^{1}P_{1}^{[1]}]\rangle|Q\bar{Q}[^{1}P_{1}^{[1]}]g\rangle\nonumber\\[1mm]
&&+\langle\mathcal
{O}^{h_{b}}[^{1}S_{0}^{[8]}]\rangle|Q\bar{Q}[^{1}S_{0}^{[8]}]g\rangle\nonumber\\[1mm]
&&+\cdots\cdots,
\end{eqnarray}
where $\langle\mathcal {O}^{H}[^{2S+1}L_{J}^{[1,8]}]\rangle$ is the
long-distance matrix element for the
bottomonium\cite{18,19,101,102,103}, and the long-distance matrix
elements for the $\Upsilon(nS)$ meson are given by
\begin{eqnarray}\label{y16}
&&\!\!\!\!\!\!\!\!\!\!\!\!\langle\mathcal
{O}^{\Upsilon(1S)}[^{3}S_{1}^{[1]}]\rangle=10.9GeV^{3},\nonumber\\[1mm]
&&\!\!\!\!\!\!\!\!\!\!\!\!\langle\mathcal
{O}^{\Upsilon(1S)}[^{1}S_{0}^{[8]}]\rangle=(0.0121\pm0.0400)GeV^{3},\nonumber\\[1mm]
&&\!\!\!\!\!\!\!\!\!\!\!\!\langle\mathcal
{O}^{\Upsilon(1S)}[^{3}S_{1}^{[8]}]\rangle=(0.0477\pm0.0334)GeV^{3},\nonumber\\[1mm]
&&\!\!\!\!\!\!\!\!\!\!\!\!\langle\mathcal
{O}^{\Upsilon(1S)}[^{3}P_{0}^{[8]}]\rangle=5\times(0.0121\pm0.0400)m_{b}^{2}\cdot
GeV^{3},
\end{eqnarray}
\begin{eqnarray}\label{y17}
&&\!\!\!\!\!\!\!\!\!\!\!\!\!\!\!\!\langle\mathcal
{O}^{\Upsilon(2S)}[^{3}S_{1}^{[1]}]\rangle=4.5GeV^{3},\nonumber\\[1mm]
&&\!\!\!\!\!\!\!\!\!\!\!\!\!\!\!\!\langle\mathcal
{O}^{\Upsilon(2S)}[^{1}S_{0}^{[8]}]\rangle=(-0.0067\pm0.0084)GeV^{3},\nonumber\\[1mm]
&&\!\!\!\!\!\!\!\!\!\!\!\!\!\!\!\!\langle\mathcal
{O}^{\Upsilon(2S)}[^{3}S_{1}^{[8]}]\rangle=(0.0224\pm0.0200)GeV^{3},\nonumber\\[1mm]
&&\!\!\!\!\!\!\!\!\!\!\!\!\!\!\!\!\langle\mathcal
{O}^{\Upsilon(2S)}[^{3}P_{0}^{[8]}]\rangle=5\times(-0.0067\pm0.0084)m_{b}^{2}\cdot
GeV^{3},
\end{eqnarray}
\begin{eqnarray}\label{y18}
&&\!\!\!\!\!\!\!\!\!\!\!\!\!\langle\mathcal
{O}^{\Upsilon(3S)}[^{3}S_{1}^{[1]}]\rangle=4.3GeV^{3},\nonumber\\[1mm]
&&\!\!\!\!\!\!\!\!\!\!\!\!\!\langle\mathcal
{O}^{\Upsilon(3S)}[^{1}S_{0}^{[8]}]\rangle=(0.0002\pm0.0062)GeV^{3},\nonumber\\[1mm]
&&\!\!\!\!\!\!\!\!\!\!\!\!\!\langle\mathcal
{O}^{\Upsilon(3S)}[^{3}S_{1}^{[8]}]\rangle=(0.0513\pm0.0085)GeV^{3},\nonumber\\[1mm]
&&\!\!\!\!\!\!\!\!\!\!\!\!\!\langle\mathcal
{O}^{\Upsilon(3S)}[^{3}P_{0}^{[8]}]\rangle=5\times(0.0002\pm0.0062)m_{b}^{2}\cdot
GeV^{3},
\end{eqnarray}

\begin{figure*}
\includegraphics[width=15cm,height=10cm]{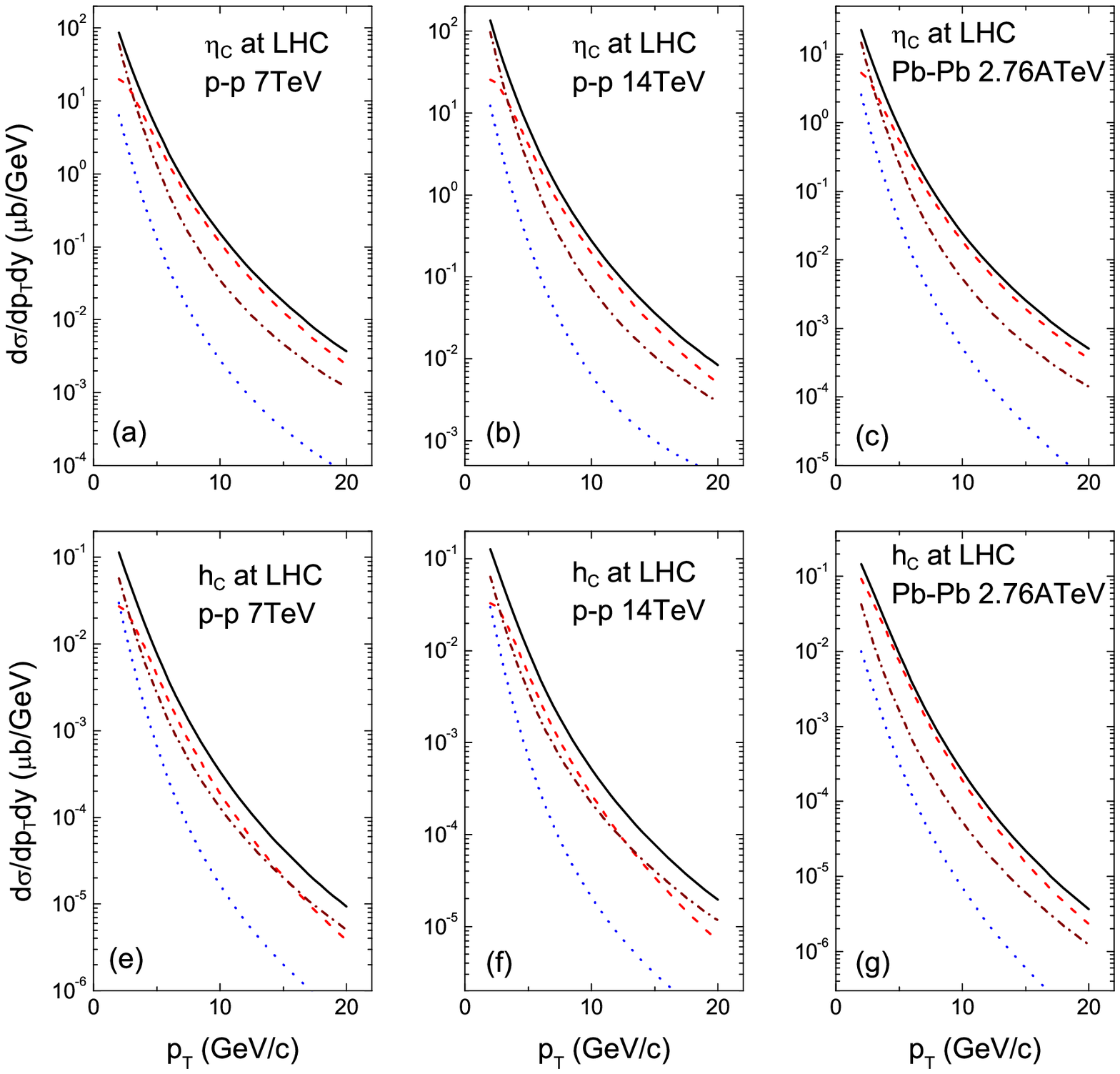}
\caption{\label{T3}  The same as Fig.\,\ref{T1} but for $\eta_{c}$
and $h_{c}$ production in p-p collisions ($\sqrt{s}=7.0 TeV$ and
$\sqrt{s}=14.0 TeV$) and Pb-Pb collisions ($\sqrt{s}=2.76 TeV$ and
$\sqrt{s}=5.5 TeV$) at LHC.}
\end{figure*}

\noindent for $\chi_{bJ}$ meson,
\begin{eqnarray}\label{y19}
&&\langle\mathcal
{O}^{\chi_{b0}(1P)}[^{3}P_{0}^{[1]}]\rangle=0.1m_{b}^{2}\cdot
GeV^{3},\nonumber\\[1mm]
&&\langle\mathcal
{O}^{\chi_{b0}(1P)}[^{3}S_{1}^{[8]}]\rangle=0.1008GeV^{3},\nonumber\\[1mm]
&&\langle\mathcal
{O}^{\chi_{b0}(2P)}[^{3}P_{0}^{[1]}]\rangle=0.036m_{b}^{2}\cdot
GeV^{3},\nonumber\\[1mm]
&&\langle\mathcal
{O}^{\chi_{b0}(2P)}[^{3}S_{1}^{[8]}]\rangle=0.0324GeV^{3},\nonumber\\[1mm]
&&\langle\mathcal
{O}^{\chi_{b1}(1P)}[^{3}P_{1}^{[1]}]\rangle=6.1GeV^{5},\nonumber\\[1mm]
&&\langle\mathcal
{O}^{\chi_{b1}(1P)}[^{3}S_{1}^{[8]}]\rangle=0.43GeV^{3},\nonumber\\[1mm]
&&\langle\mathcal
{O}^{\chi_{b1}(2P)}[^{3}P_{1}^{[1]}]\rangle=7.1m_{b}^{2}GeV^{3},\nonumber\\[1mm]
&&\langle\mathcal
{O}^{\chi_{b1}(2P)}[^{3}S_{1}^{[8]}]\rangle=0.52GeV^{3},\nonumber\\[1mm]
&&\langle\mathcal
{O}^{\chi_{b1}(3P)}[^{3}P_{1}^{[1]}]\rangle=7.7m_{b}^{2}GeV^{3},
\end{eqnarray}
for $\eta_{b}$ meson,
\begin{eqnarray}\label{y20}
&&\langle\mathcal
{O}^{\eta_{b}(nS)}[^{1}S_{0}^{[1]}]\rangle=\frac{1}{3}\times\langle\mathcal
{O}^{\Upsilon(nS)}[^{3}S_{1}^{[1]}]\rangle,\nonumber\\[1mm]
&&\langle\mathcal
{O}^{\eta_{b}(nS)}[^{1}S_{0}^{[8]}]\rangle=\frac{1}{3}\times\langle\mathcal
{O}^{\Upsilon(nS)}[^{3}S_{1}^{[8]}]\rangle,\nonumber\\[1mm]
&&\langle\mathcal
{O}^{\eta_{b}(nS)}[^{3}S_{1}^{[8]}]\rangle=\langle\mathcal
{O}^{\Upsilon(nS)}[^{1}S_{0}^{[8]}]\rangle,\nonumber\\[1mm]
&&\langle\mathcal
{O}^{\eta_{b}(nS)}[^{1}P_{1}^{[8]}]\rangle=3\times\langle\mathcal
{O}^{\Upsilon(nS)}[^{3}P_{0}^{[8]}]\rangle,
\end{eqnarray}
and for $h_{b}$ meson,
\begin{eqnarray}\label{y21}
&&\langle\mathcal
{O}^{h_{b}(nP)}[^{1}P_{1}^{[1]}]\rangle=3\times\langle\mathcal
{O}^{\chi_{b0}(nP)}[^{3}P_{0}^{[1]}]\rangle,\nonumber\\[1mm]
&&\langle\mathcal
{O}^{h_{b}(nP)}[^{1}S_{0}^{[8]}]\rangle=3\times\langle\mathcal
{O}^{\chi_{b0}(nP)}[^{3}S_{1}^{[8]}]\rangle,
\end{eqnarray}
here $m_{b}$ is the bottom quark mass.

\begin{figure*}
\includegraphics[width=15cm,height=10cm]{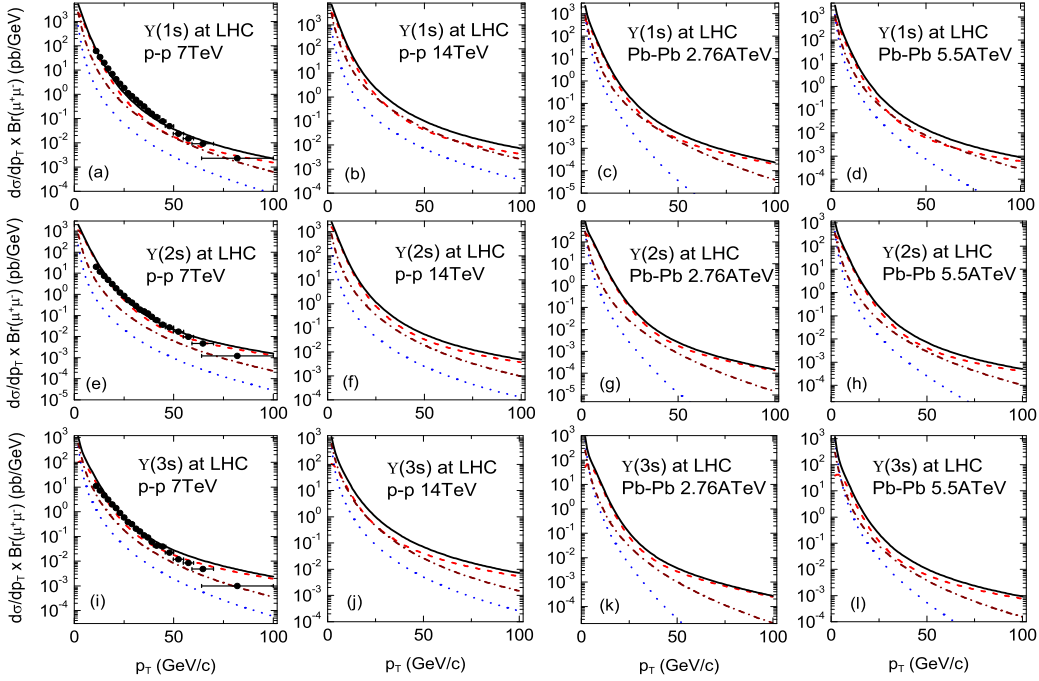}
\caption{\label{T4}  The invariant cross section of large-$p_{T}$
$\Upsilon(nS)$ production for p-p collisions ($\sqrt{s}=7.0 TeV$ and
$\sqrt{s}=14.0 TeV$) and Pb-Pb collisions ($\sqrt{s}=2.76 TeV$ and
$\sqrt{s}=5.5 TeV$) at LHC. The dashed line (red line) for the
initial parton hard scattering processes, the dotted line (blue
line) for the semielastic photoproduction processes, the
dashed-dotted line (wine line) for the inelastic photoproduction
processes. The branching fractions are
$B_{r}(\Upsilon(1S)\rightarrow\mu^{+}\mu^{-})=2.48\%$,
$B_{r}(\Upsilon(2S)\rightarrow\mu^{+}\mu^{-})=1.93\%$, and
$B_{r}(\Upsilon(3S)\rightarrow\mu^{+}\mu^{-})=2.18\%$. The
$\Upsilon(nS)$ data points are from the CMS Collaboration\cite{9}.}
\end{figure*}

In the non-relativistic Quantum Chromodynamics (NRQCD) factorization
approach, the hadroproduction cross section for the heavy quarkonium
produced by the initial parton hard scattering processes (LO) can be
expressed as
\begin{eqnarray}\label{y22}
&&\frac{d\sigma_{AB\rightarrow
HX}^{LO}}{dp_{T}^{2}dy}=\sum_{n}\frac{d\sigma_{AB\rightarrow
Q\bar{Q}_{[1,8]}(n)+X}^{LO}}{dp_{T}^{2}dy}\langle\mathcal
{O}_{[1,8]}^{H}[n]\rangle\nonumber\\[1mm]
&&=\sum_{n}\int
dx_{a}f_{A}(x_{a},Q^{2})f_{B}(x_{b},Q^{2})\frac{x_{a}x_{b}}{x_{a}-x_{1}}\nonumber\\[1mm]
&&\times\frac{d\hat{\sigma}}{d\hat{t}}(a+b\rightarrow
Q\bar{Q}_{[1,8]}[n]+X)\langle\mathcal {O}_{[1,8]}^{H}[n]\rangle,
\end{eqnarray}
where $x_{a}$ and $x_{b}=(x_{a}x_{2}-\tau)/(x_{a}-x_{1})$ are the
momentum fractions of the partons. The variables are
$x_{1}=\frac{1}{2}(x_{T}^{2}+4\tau)^{1/2}\exp(y)$,
$x_{2}=\frac{1}{2}(x_{T}^{2}+4\tau)^{1/2}\exp(-y)$,
$x_{T}=2p_{T}/\sqrt{s}$, $\tau=M^{2}/s$, and $M$ is the mass of the
heavy quarkonium.

\begin{figure*}
\includegraphics[width=15cm,height=10cm]{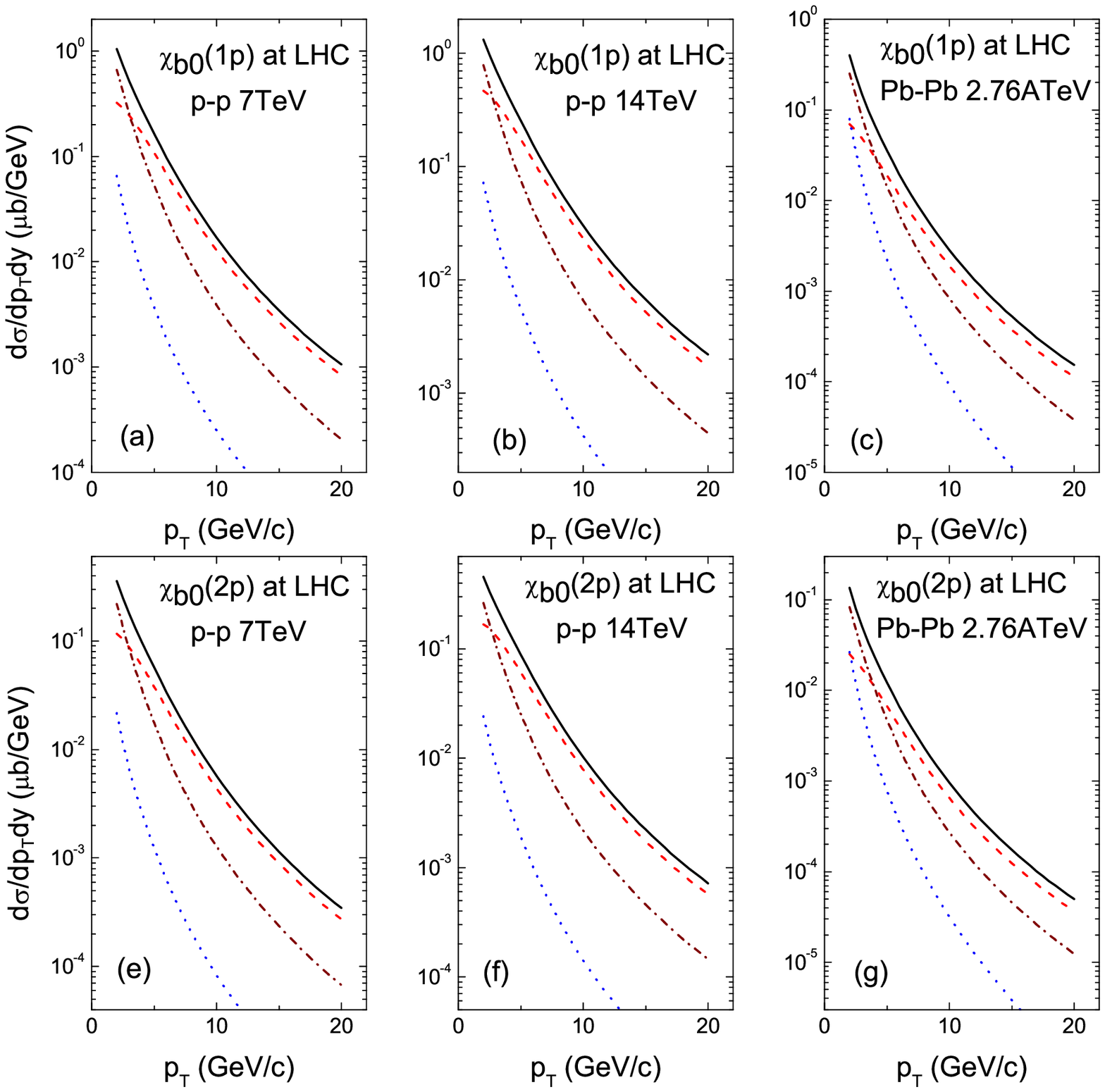}
\caption{\label{T5}  The same as Fig.\,\ref{T4} but for $\chi_{b0}$
production in p-p collisions ($\sqrt{s}=7.0 TeV$ and $\sqrt{s}=14.0
TeV$) and Pb-Pb collisions ($\sqrt{s}=2.76 TeV$ and $\sqrt{s}=5.5
TeV$) at LHC.}
\end{figure*}

The parton distribution function $f_{A}(x,Q^{2})$ of the nucleon is
given by\cite{104,105,106,107}
\begin{eqnarray}\label{y23}
f_{A}(x,Q^{2})=R_{A}(x,Q^{2})[\frac{Z}{A}p(x,Q^{2})+\frac{N}{A}n(x,Q^{2})],
\end{eqnarray}
where $R_{A}(x,Q^{2})$ is the nuclear modification factor\cite{108},
$Z$ is the proton number, $N$ is the neutron number, and $A$ is the
nucleon number; $p(x,Q^{2})$ and $n(x,Q^{2})$ are the parton
distribution functions of protons and neutrons, respectively.

The differential cross section of the initiated partonic
subprocesses was calculated in Ref.\cite{77} and can be written as
\begin{eqnarray}\label{y24}
&&\!\!\!\!\frac{d\hat{\sigma}}{d\hat{t}}(a+b\rightarrow H+X)\nonumber\\[1mm]
&&\!\!\!\!=\frac{d\hat{\sigma}}{d\hat{t}}(a+b\rightarrow
Q\bar{Q}_{[1,8]}[n]+X)\langle\mathcal {O}_{[1,8]}^{H}[n]\rangle\nonumber\\[1mm]
&&\!\!\!\!=\frac{1}{16\pi\hat{s}^{2}}\left|M[^{2S+1}L_{J}^{[1,8]}]\right|^{2}\frac{\langle0|\mathcal
{O}^{H}[^{2S+1}L_{J}^{[1,8]}]|0\rangle}{m_{Q}(2J+1)},
\end{eqnarray}
where $m_{Q}$ is the mass of the heavy flavor quark.

The fragmentation cross section for the heavy quarkonium produced by
the initial parton hard scattering processes (LO-fra.) can be
expressed as
\begin{eqnarray}\label{y25}
&&\frac{d\sigma_{AB\rightarrow
HX}^{LO-fra.}}{dp_{T}^{2}dy}=\sum_{n}\frac{d\sigma_{AB\rightarrow
Q\bar{Q}_{[1,8]}(n)+X}^{LO-fra.}}{dp_{T}^{2}dy}\langle\mathcal
{O}_{[1,8]}^{H}[n]\rangle\nonumber\\[1mm]
&&=\sum_{n}\int
dx_{a}dx_{b}f_{A}(x_{a},Q^{2})f_{B}(x_{b},Q^{2})\frac{x_{a}x_{b}}{z_{c}(x_{a}x_{b}-\tau)}\nonumber\\[1mm]
&&\times\frac{d\hat{\sigma}}{d\hat{t}}(ab\rightarrow
cd)D_{c\rightarrow
Q\bar{Q}_{[1,8]}[n]}\left(z_{c},Q^{2}\right)\langle\mathcal
{O}_{[1,8]}^{H}[n]\rangle,
\end{eqnarray}
where $z_{c}$ is the momentum fraction of the final heavy
quarkonium, $\frac{d\hat{\sigma}}{d\hat{t}}(ab\rightarrow cd)$ is
the differential cross section for the subprocess\cite{109}, and the
fragmentation function $D_{c\rightarrow
Q\bar{Q}_{[1,8]}[n]}(z_{c},Q^{2})$ was calculated in
Ref.\cite{66,67,68,69,70}.

The large-$p_{T}$ heavy quarkonium produced by semielastic
photoproduction processes can be divided into the semielastic direct
photoproduction processes and semielastic resolved photoproduction
processes in ultrarelativistic heavy ion collisions.

In the semielastic direct photoproduction processes, incident
nucleus can emit a photon, then the high-energy photon interacts
with parton of another incident nucleus by the interaction of
quark-photon Compton scattering and gluon-photon fusion. The
differential cross section of heavy quarkonium produced by the
semielastic direct photoproduction processes (semi.dir. and
semi.dir.-fra.) can be written as
\begin{eqnarray}\label{y26}
&&\frac{d\sigma_{AB\rightarrow
HX}^{semi.dir.}}{dp_{T}^{2}dy}=\sum_{n}\frac{d\sigma_{AB\rightarrow
Q\bar{Q}_{[1,8]}(n)+X}^{semi.dir.}}{dp_{T}^{2}dy}\langle\mathcal
{O}_{[1,8]}^{H}[n]\rangle\nonumber\\[1mm]
&&=\sum_{n}\int
dx_{a}f_{\gamma/N}(x_{a})f_{B}(x_{b},Q^{2})\frac{x_{a}x_{b}}{x_{a}-x_{1}}\nonumber\\[1mm]
&&\times\frac{d\hat{\sigma}}{d\hat{t}}(\gamma+b\rightarrow
Q\bar{Q}_{[1,8]}[n]+X)\langle\mathcal {O}_{[1,8]}^{H}[n]\rangle,
\end{eqnarray}
\begin{eqnarray}\label{y27}
&&\frac{d\sigma_{AB\rightarrow
HX}^{semi.dir.-fra.}}{dp_{T}^{2}dy}=\sum_{n}\frac{d\sigma_{AB\rightarrow
Q\bar{Q}_{[1,8]}(n)+X}^{semi.dir.-fra.}}{dp_{T}^{2}dy}\langle\mathcal
{O}_{[1,8]}^{H}[n]\rangle\nonumber\\[1mm]
&&=\sum_{n}\int
dx_{a}dx_{b}f_{\gamma/N}(x_{a})f_{B}(x_{b},Q^{2})\frac{x_{a}x_{b}}{z_{c}(x_{a}x_{b}-\tau)}\nonumber\\[1mm]
&&\times\frac{d\hat{\sigma}}{d\hat{t}}(\gamma b\rightarrow
cd)D_{c\rightarrow
Q\bar{Q}_{[1,8]}[n]}\left(z_{c},Q^{2}\right)\langle\mathcal
{O}_{[1,8]}^{H}[n]\rangle,
\end{eqnarray}
Here the differential cross section of the subprocesses was
calculated in Ref.\cite{77}.

The equivalent photon spectrum for nucleus can be obtained from the
semiclassical description of high-energy electromagnetic collisions.
A relativistic nucleus with $Z$ times the electric charge moving
with a relativistic factor $\gamma\gg1$ with respect to some
observers develop an equally strong magnetic field component so it
resembles a beam of real photons, where the number of photons can be
expressed as\cite{110,111}
\begin{eqnarray}\label{y28}
f_{\gamma/N}(\omega)=\frac{2Z^{2}\alpha}{\pi\omega}\ln\left(\frac{\gamma}{\omega
R}\right),
\end{eqnarray}
where $ω$ is the photon momentum, and $R=b_{min}$ is the radius of
the nucleus ($b_{min}$ is the cutoff of impact parameter). In the
logarithmic approximation, the results obtained from the purely
classical treatment or by including the form factor are related to
each other through the rescaling of the relativistic factor.

\begin{figure*}
\includegraphics[width=15cm,height=10cm]{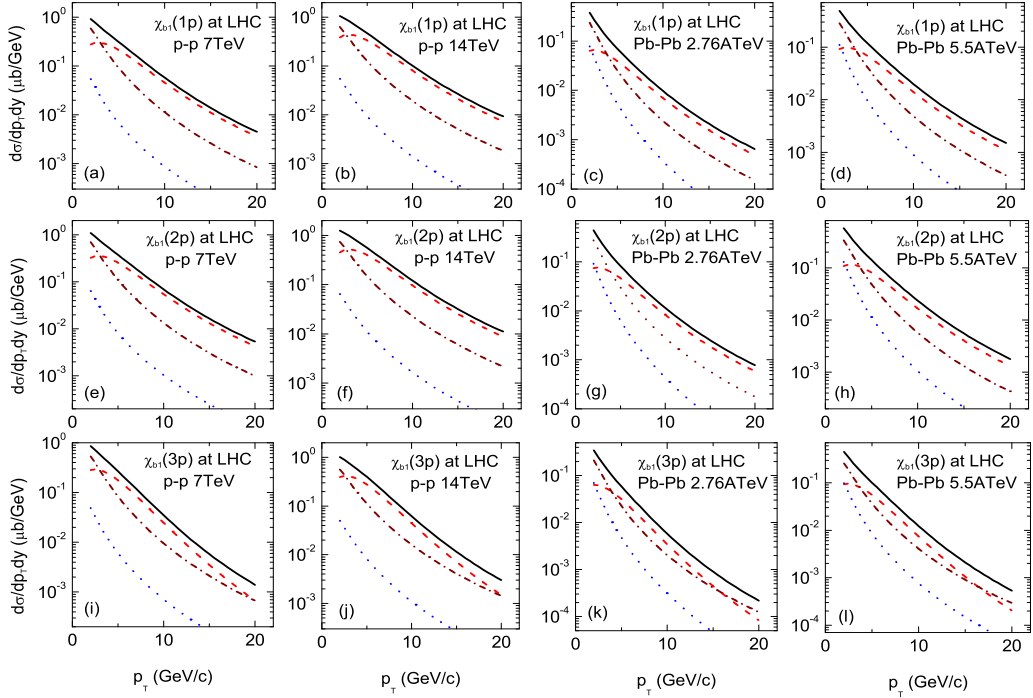}
\caption{\label{T6}  The same as Fig.\,\ref{T4} but for $\chi_{b1}$
production in p-p collisions ($\sqrt{s}=7.0 TeV$ and $\sqrt{s}=14.0
TeV$) and Pb-Pb collisions ($\sqrt{s}=2.76 TeV$ and $\sqrt{s}=5.5
TeV$) at LHC.}
\end{figure*}

For p-p collisions, the equivalent photon spectrum function for the
proton can be obtained from the Weizs$\ddot{a}$cker-Williams
approximation\cite{112,113,114},
\begin{eqnarray}\label{y29}
f_{\gamma/p}(x)&=&\frac{\alpha}{2\pi}\frac{1+(1-x)^{2}}{x}\nonumber\\[1mm]
&&\times\left[\ln
A_{p}-\frac{11}{6}+\frac{3}{A_{p}}-\frac{3}{2A_{p}^{2}}+\frac{1}{3A_{p}^{3}}\right],
\end{eqnarray}
where $x$ is the momentum fraction of the photon,
$A_{p}=1+0.71GeV^{2}/Q_{min}^{2}$ with
\begin{eqnarray}\label{y30}
Q_{min}^{2}&=&-2m_{p}^{2}+\frac{1}{2s}\bigg\{(s+m_{p}^{2})(s-xs+m_{p}^{2})\nonumber\\[1mm]
&&-(s-m_{p}^{2})\sqrt{(s-xs+m_{p}^{2})^{2}-4m_{p}^{2}xs}\bigg\},\,
\end{eqnarray}
here $m_{p}$ is the the mass of the proton, and at high energies
$Q_{min}^{2}$ is given to a very good approximation by
$m_{p}^{2}x^{2}/(1-x)$. Propagated uncertainties to the final cross
sections are on the order of 10\% for p-p collisions and 20\% for
Pb-Pb collisions since covering different form-factor
parametrizations and the convolution of the nuclear photon
fluxes\cite{115}.

In the semielastic resolved photoproduction processes, the parton
from the hadron-like photon emitted by incident nucleus can interact
with the parton of another incident nucleus via the interactions of
quark-antiquark annihilation, quark-gluon Compton scattering, and
gluon-gluon fusion. The invariant cross section for large-$p_{T}$
heavy quarkonium produced by the semielastic resolved
photoproduction processes (semi.res. and semi.res.-fra.) in the
hadronic collisions can be written as
\begin{eqnarray}\label{y31}
&&\!\!\!\!\!\!\!\!\!\!\!\!\!\frac{d\sigma_{AB\rightarrow
HX}^{semi.res.}}{dp_{T}^{2}dy}=\sum_{n}\frac{d\sigma_{AB\rightarrow
Q\bar{Q}_{[1,8]}(n)+X}^{semi.res.}}{dp_{T}^{2}dy}\langle\mathcal
{O}_{[1,8]}^{H}[n]\rangle\nonumber\\[1mm]
&&\!\!\!\!\!\!\!\!\!\!\!\!\!=\sum_{n}\int
dx_{a}dx_{b}f_{\gamma/N}(x_{a})f_{\gamma}(z_{a},Q^{2})f_{B}(x_{b},Q^{2})\nonumber\\[1mm]
&&\!\!\!\!\!\!\!\!\!\!\!\!\!\times\frac{x_{a}x_{b}z_{a}}{x_{a}x_{b}-x_{a}x_{2}}\frac{d\hat{\sigma}}{d\hat{t}}(a'b\rightarrow
Q\bar{Q}_{[1,8]}[n]+X)\langle\mathcal {O}_{[1,8]}^{H}[n]\rangle,
\end{eqnarray}
\begin{eqnarray}\label{y32}
&&\!\!\!\!\!\!\!\!\frac{d\sigma_{AB\rightarrow
HX}^{semi.res.-fra.}}{dp_{T}^{2}dy}=\sum_{n}\frac{d\sigma_{AB\rightarrow
Q\bar{Q}_{[1,8]}(n)+X}^{semi.res.-fra.}}{dp_{T}^{2}dy}\langle\mathcal
{O}_{[1,8]}^{H}[n]\rangle\nonumber\\[1mm]
&&\!\!\!\!\!\!\!\!=\sum_{n}\int
dx_{a}dx_{b}dz_{a}f_{\gamma/N}(x_{a})f_{\gamma}(z_{a},Q^{2})f_{B}(x_{b},Q^{2})\nonumber\\[1mm]
&&\!\!\!\!\!\!\!\!\times\frac{x_{a}x_{b}z_{a}}{z_{c}(x_{a}x_{b}z_{a}\!-\!\!\tau)}\frac{d\hat{\sigma}}{d\hat{t}}(a'\!b\!\rightarrow\!cd)D_{\!c\rightarrow
Q\bar{Q}_{[1,8]}\![n]}\!\!\left(z_{c},Q^{2}\!\right)\!\!\langle\mathcal
{O}_{[1,8]}^{H}\![n]\rangle,\nonumber\\[1mm]
\end{eqnarray}
where $f_{\gamma}(z_{a},Q^{2})$ is the parton distribution function
of the hadron-like photon\cite{116}, the differential cross section
of the subprocesses was calculated in Ref.\cite{77}.

The large-$p_{T}$ heavy quarkonium produced by inelastic
photoproduction processes can be divided into the inelastic direct
photoproduction processes and inelastic resolved photoproduction
processes in ultrarelativistic heavy ion collisions.

\begin{figure*}
\includegraphics[width=15cm,height=10cm]{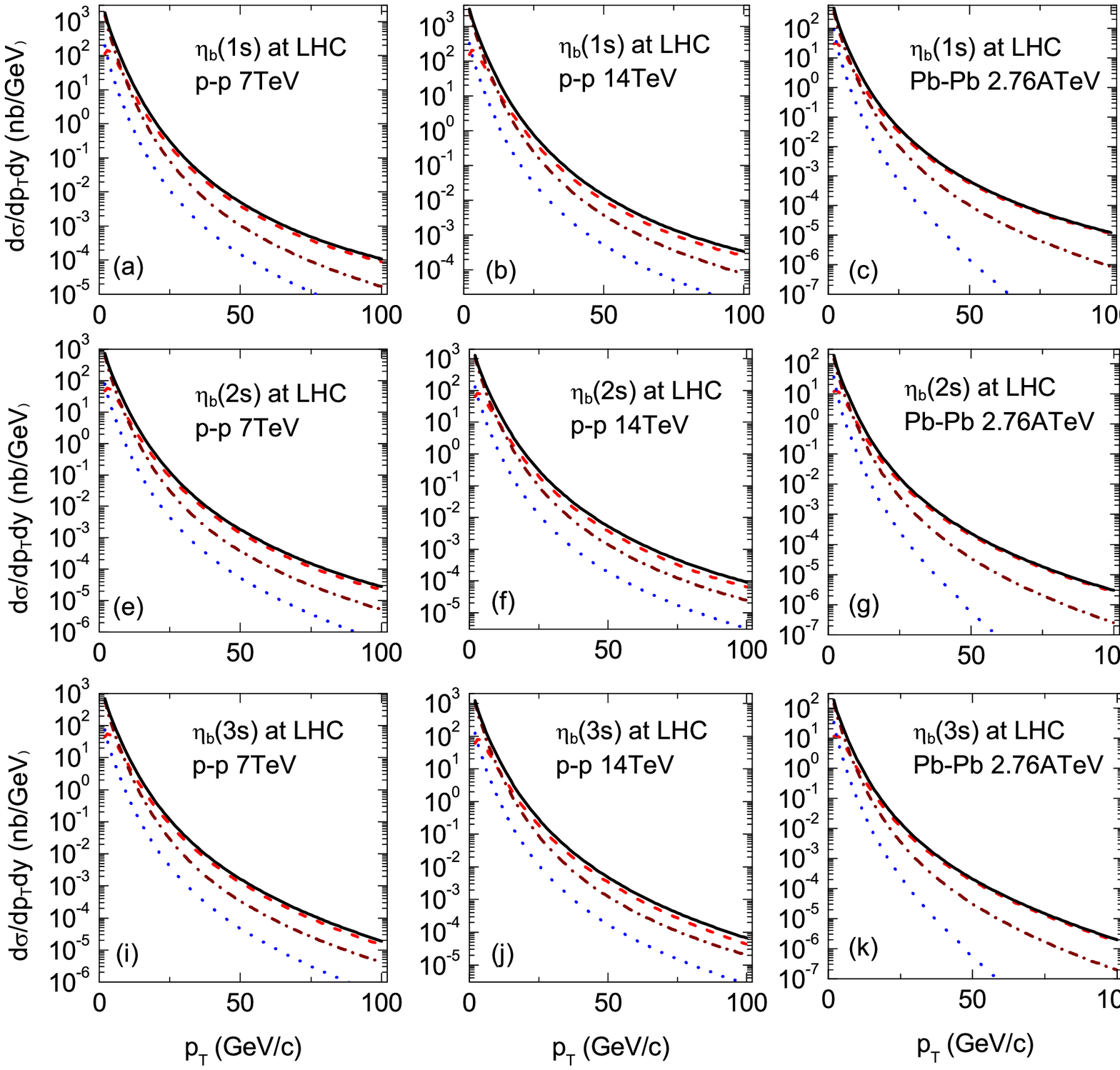}
\caption{\label{T7}  The same as Fig.\,\ref{T4} but for $\eta_{b}$
production in p-p collisions ($\sqrt{s}=7.0 TeV$ and $\sqrt{s}=14.0
TeV$) and Pb-Pb collisions ($\sqrt{s}=2.76 TeV$ and $\sqrt{s}=5.5
TeV$) at LHC.}
\end{figure*}

In the inelastic direct photoproduction processes, the charged
parton of the incident nucleus can emit a photon, then the high
energy photon interacts with parton of another incident nucleus by
the interaction of quark-photon Compton scattering and gluon-photon
fusion. The differential cross section of heavy quarkonium produced
by the inelastic direct photoproduction processes (inel.dir. and
inel.dir.-fra.) in the hadronic collisions can be expressed as
\begin{eqnarray}\label{y33}
&&\!\!\!\!\!\!\!\!\!\!\!\!\frac{d\sigma_{AB\rightarrow
HX}^{inel.dir.}}{dp_{T}^{2}dy}=\sum_{n}\frac{d\sigma_{AB\rightarrow
Q\bar{Q}_{[1,8]}(n)+X}^{inel.dir.}}{dp_{T}^{2}dy}\langle\mathcal
{O}_{[1,8]}^{H}[n]\rangle\nonumber\\[1mm]
&&\!\!\!\!\!\!\!\!\!\!\!\!=\sum_{n}\int
dx_{a}dx_{b}f_{A}(x_{a},Q^{2})f_{\gamma/q}(z_{a})f_{B}(x_{b},Q^{2})\nonumber\\[1mm]
&&\!\!\!\!\!\!\!\!\!\!\!\!\times\frac{x_{a}x_{b}z_{a}}{x_{a}x_{b}-x_{a}x_{2}}\frac{d\hat{\sigma}}{d\hat{t}}(\gamma
b\rightarrow Q\bar{Q}_{[1,8]}[n]+X)\langle\mathcal
{O}_{[1,8]}^{H}[n]\rangle,
\end{eqnarray}
\begin{eqnarray}\label{y34}
&&\!\!\!\!\!\!\!\!\frac{d\sigma_{AB\rightarrow
HX}^{inel.dir.-fra.}}{dp_{T}^{2}dy}=\sum_{n}\frac{d\sigma_{AB\rightarrow
Q\bar{Q}_{[1,8]}(n)+X}^{inel.dir.-fra.}}{dp_{T}^{2}dy}\langle\mathcal
{O}_{[1,8]}^{H}[n]\rangle\nonumber\\[1mm]
&&\!\!\!\!\!\!\!\!=\sum_{n}\int
dx_{a}dx_{b}dz_{a}f_{A}(x_{a},Q^{2})f_{\gamma/q}(z_{a})f_{B}(x_{b},Q^{2})\nonumber\\[1mm]
&&\!\!\!\!\!\!\!\!\times\frac{x_{a}x_{b}z_{a}}{z_{c}(x_{a}x_{b}z_{a}\!-\!\!\tau)}\frac{d\hat{\sigma}}{d\hat{t}}(\gamma\!b\!\rightarrow\!cd)D_{\!c\rightarrow
Q\bar{Q}_{[1,8]}\![n]}\!\!\left(z_{c},Q^{2}\!\right)\!\!\langle\mathcal
{O}_{[1,8]}^{H}\![n]\rangle,\nonumber\\[1mm]
\end{eqnarray}
here the equivalent photon spectrum function of the charged parton
is given by\cite{117,118}
\begin{eqnarray}\label{y35}
f_{\gamma/q}(x)&=&\frac{\alpha}{2\pi}e_{f}^{2}\bigg\{\frac{1+(1-x)^{2}}{x}\left[\ln\left(\frac{E}{m}\right)-\frac{1}{2}\right]\nonumber\\[1mm]
&&+\!\frac{x}{2}\left[\ln\left(\frac{2}{x}\!-\!2\right)\!+\!1\right]\!+\!\frac{(2\!-\!x)^{2}}{2x}\ln\left(\frac{2\!-\!2x}{2\!-\!x}\right)\bigg\},\nonumber\\[1mm]
\end{eqnarray}
where $x$ is the the photon momentum fraction. The variables
$e_{f}$, $E$, and $m$ are the charge, energy, and mass of the
parton, respectively.

In the inelastic resolved photoproduction processes, the parton from
the hadron-like photon emitted by the charged parton of incident
nucleus can interact with the parton of another incident nucleus via
the interactions of quark-antiquark annihilation, quark-gluon
Compton scattering, and gluon-gluon fusion. The invariant cross
section for large-$p_{T}$ heavy quarkonium produced by the inelastic
resolved photoproduction processes (inel.res. and inel.res.-fra.) in
the hadronic collisions can be written as
\begin{eqnarray}\label{y36}
&&\!\!\!\!\!\!\!\!\!\frac{d\sigma_{AB\rightarrow
HX}^{inel.res.}}{dp_{T}^{2}dy}=\sum_{n}\frac{d\sigma_{AB\rightarrow
Q\bar{Q}_{[1,8]}(n)+X}^{inel.dir.}}{dp_{T}^{2}dy}\langle\mathcal
{O}_{[1,8]}^{H}[n]\rangle\nonumber\\[1mm]
&&\!\!\!\!\!\!\!\!\!=\!\sum_{n}\!\int\!
dx_{a}dx_{b}dz_{a}f_{A}(x_{a},\!Q^{2})f_{\gamma/q}(z_{a})f_{\gamma}(z'_{a},\!Q^{2})f_{B}(x_{b},\!Q^{2})\nonumber\\[1mm]
&&\!\!\!\!\!\!\!\!\!\times\frac{x_{a}x_{b}z_{a}z'_{a}}{x_{a}x_{b}z_{a}\!-\!x_{a}z_{a}x_{2}}\frac{d\hat{\sigma}}{d\hat{t}}(a'
b\rightarrow\!\! Q\bar{Q}_{[1,8]}[n]\!+\!X)\langle\mathcal
{O}_{[1,8]}^{H}[n]\rangle,
\end{eqnarray}

\begin{figure*}
\includegraphics[width=15cm,height=10cm]{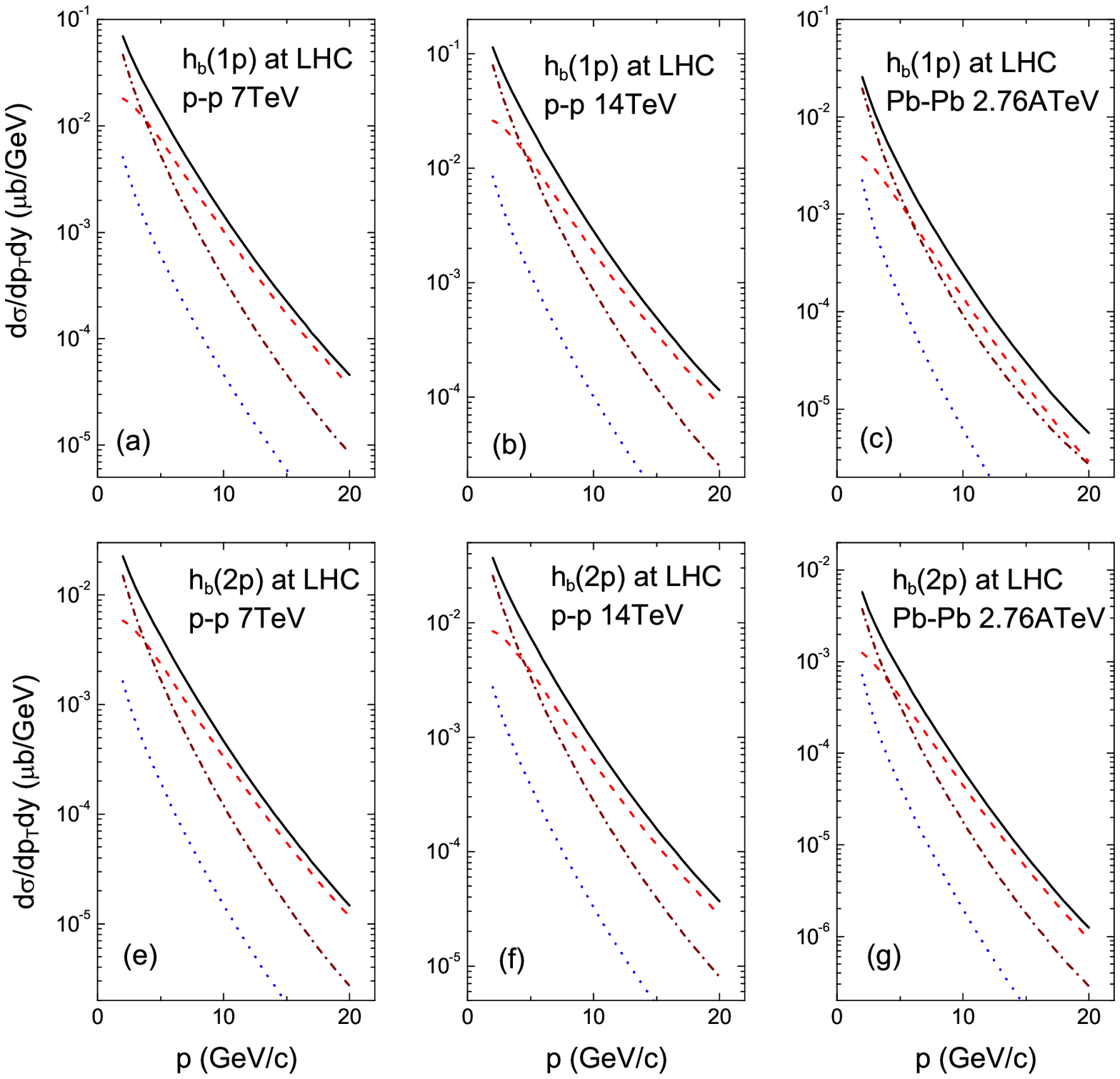}
\caption{\label{T8}  The same as Fig.\,\ref{T4} but for $h_{b}$
production in p-p collisions ($\sqrt{s}=7.0 TeV$ and $\sqrt{s}=14.0
TeV$) and Pb-Pb collisions ($\sqrt{s}=2.76 TeV$ and $\sqrt{s}=5.5
TeV$) at LHC.}
\end{figure*}

\begin{eqnarray}\label{y37}
&&\!\!\!\!\!\!\!\!\!\!\!\!\frac{d\sigma_{AB\rightarrow
HX}^{inel.res.-fra.}}{dp_{T}^{2}dy}=\sum_{n}\frac{d\sigma_{AB\rightarrow
Q\bar{Q}_{[1,8]}(n)+X}^{inel.res.-fra.}}{dp_{T}^{2}dy}\langle\mathcal
{O}_{[1,8]}^{H}[n]\rangle\nonumber\\[1mm]
&&\!\!\!\!\!\!\!\!\!\!\!\!=\!\!\sum_{n}\!\!\int\!\!\!
dx_{a}dx_{b}dz_{a}dz'_{a}f_{\!A}\!(x_{a},\!Q^{2})f_{\!\gamma\!/q}(z_{a})f_{\!\gamma}\!(z'_{a},\!Q^{2})f_{\!B}\!(x_{b},\!Q^{2})\nonumber\\[1mm]
&&\!\!\!\!\!\!\!\!\!\!\!\!\!\!\times\frac{x_{a}x_{b}z_{a}z'_{a}}{z_{c}(x_{a}x_{b}z_{a}z'_{a}\!-\!\!\tau)}\frac{d\hat{\sigma}}{d\hat{t}}(a'b\!\rightarrow\!cd)D_{\!c\rightarrow
Q\bar{Q}_{[1,8]}\![n]}\!\!\left(z_{c},\!Q^{2}\!\right)\!\!\langle\mathcal
{O}_{[1,8]}^{H}\![n]\rangle,\nonumber\\[1mm]
\end{eqnarray}
where $z'_{a}$ is the momentum fraction of the parton from the
hadron-like photon.


\section{Numerical results}

In ultrarelativistic heavy ion collisions, the equivalent photon
spectrum for the nucleus obtained from semiclassical description of
high energies electromagnetic collisions is $f_{\gamma/N}\propto
Z^{2}\ln\gamma$, the relativistic factor
$\gamma=E/m_{N}=\sqrt{s}/2m_{N}\gg1$ becomes very large at Large
Hadron Collider energies. Indeed, the equivalent photon spectrum
function for the proton obtained from Weizs$\ddot{a}$cker-Williams
approximation is $f_{\gamma/p}\propto\ln A\propto\ln(s/m_{p}^{2})$,
where $m_{p}$ is the proton mass. Since the collision energy
$\sqrt{s}$ at Large Hadron Collider is very large, the photon
spectrum becomes important. Therefore the contribution of
semielastic photoproduction processes is evident at Large Hadron
Collider energies. For the inelastic photoproduction processes, the
equivalent photon spectrum function of the charged parton is
$f_{\gamma/q}\propto\ln(E/m_{q})=\ln(\sqrt{s}/2m_{q})+\ln(x)$, where
$m_{q}$ is the charged parton mass. Hence the photon spectrum for
the charged parton becomes prominent at Large Hadron Collider
energies. Therefore the contribution of photoproduction processes is
evident at LHC. Moreover, we also calculate the fragmentation contribution of heavy quarkonium.
For inclusive heavy quarkonium production, there is a large contribution from the $g\rightarrow H$ fragmentation, where $H$ is the heavy quarkonium. At large
transverse momentum, this is comparable to the direct production processes over almost the whole range of transverse momentum values considered.
The numerical results of our calculation for large-$p_{T}$
heavy quarkonium produced by the hard photoproduction processes in
relativistic heavy ion collisions are plotted in
Figs.\,\ref{T1}-\ref{T8}.

In Figs.\,\ref{T1}-\ref{T3}, we plot the contribution of the
charmonium ($J/\psi$, $\psi(2s)$, $\chi_{cJ}$, $\eta_{c}$, and
$h_{c}$) produced by the semielastic and inelastic hard
photoproduction (and fragmentation) processes for p-p collisions ($\sqrt{s}=7 TeV$ and
$\sqrt{s}=14 TeV$) and Pb-Pb collisions ($\sqrt{s}=2.76 TeV$ and
$\sqrt{s}=5.5 TeV$) in relativistic heavy ion collisions. The
charmonium spectra of semielastic and inelastic photoproduction (and fragmentation)
processes (the dotted line and dashed-dotted line) are compared with
the charmonium spectra of the initial parton hard scattering
processes (the dashed line) for p-p collisions and Pb-Pb collisions
at the LHC, respectively. In panel (a) and (e) of Figs.\,\ref{T1},
we compare with the ALICE Collaboration $J/\psi$ meson data\cite{1}
and the CMS Collaboration $\psi(2S)$ meson data\cite{8} in p-p
collisions with $\sqrt{s}=7 TeV$. We find that the contribution of
$J/\psi$ and $\psi(2S)$ meson produced by hard photoproduction and fragmentation
processes cannot be negligible for large-$p_{T}$ in p-p collisions
with $\sqrt{s}=7 TeV$ at LHC [Figs.\,\ref{T1}(a) and (e)]. We also
plot the spectra of bottomonium ($\Upsilon(ns)$, $\chi_{bJ}$,
$\eta_{b}$, and $h_{b}$) produced by the hard photoproduction and fragmentation
processes for p-p collisions ($\sqrt{s}=7 TeV$ and $\sqrt{s}=14
TeV$) and Pb-Pb collisions ($\sqrt{s}=2.76 TeV$ and $\sqrt{s}=5.5
TeV$) in Figs.\,\ref{T4}-\ref{T8}. Compared with the production
cross section of the initial parton hard scattering processes (the
dashed line), the contribution of the bottomonium produced by
semielastic and inelastic hard photoproduction processes (the dotted
line and dashed-dotted line) is prominent at LHC energies.
Furthermore, we compare with the CMS Collaboration $\Upsilon(nS)$
meson data\cite{9} for p-p collisions with $\sqrt{s}=7 TeV$ in panel
(a), (e), and (i) of Figs.\,\ref{T4}. We find that the contribution
of $\Upsilon(nS)$ meson produced by photoproduction processes and fragmentation
becomes is evident for p-p collisions with $\sqrt{s}=7 TeV$ at
LHC[Figs.\,\ref{T4}(a), (e), and (i)].


\section{Conclusion}

In the framework of the non-relativistic Quantum Chromodynamics
(NRQCD) formalism, we have investigated the production of heavy
quarkonium by the semielastic hard photoproduction
processes, inelastic hard photoproduction
processes and fragmentation processes in p-p collisions and Pb-Pb collisions in
ultrarelativistic heavy ion collisions. At the early stage of
relativistic heavy ion collisions, the incident nucleus (the charged
parton of the incident nucleus) can emit large-$p_{T}$ photons then
the high energy photons interact with the partons of another
incident nucleon by the quark-photon Compton scattering and
photon-gluon fusion interactions. Furthermore, the partons of the
hadron-like photons can interact with the partons of the nucleus by
the quarkantiquark annihilation and quark-gluon Compton scattering
interactions. Indeed, we also calculate the fragmentation contribution of heavy quarkonium.
For inclusive heavy quarkonium production, the contribution of $g\rightarrow H$ fragmentation processes
is comparable to the direct production processes over almost the whole range of transverse momentum values considered.
The numerical results indicate that the contribution of heavy quarkonium produced by the hard photoproduction processes and fragmentation processes
becomes evident in p-p collisions ($\sqrt{s}=7 TeV$ and $\sqrt{s}=14.0
TeV$) and Pb-Pb collisions ($\sqrt{s}=2.76 TeV$ and $\sqrt{s}=5.5
TeV$) at Large Hadron Collider (LHC) energies.


\section{Acknowledgements}

This work is supported by the National Basic Research Program of China (973 Program) with Grant No.
2014CB845405 and the National Natural Science Foundation of
China with Grant No. 11465021 and No. 11065010.







\bibliographystyle{model1-num-names}
\bibliography{<your-bib-database>}

\end{document}